\documentclass{emulateapj}

\usepackage{graphicx}
\usepackage{lscape}
\usepackage{natbib}
\def\HII{{\ion{H}{2}}}

\def\OII{[{\ion{O}{2}}]}

\def\OIII{[{\ion{O}{3}}]}
\def\OIII5007Hb{[{\ion{O}{3}}]~$\lambda5007$/H$\beta$}
\def\4959_5007{[\ion{O}{3}]~$\lambda \lambda$4959,5007}
\def\OIII49595007{[\ion{O}{3}]~$\lambda \lambda 4959,5007$}
\def\ratioR23{([\ion{O}{2}]~$\lambda$3727 +
[\ion{O}{3}]~$\lambda\lambda$4959,5007)/H$\beta$}
\def\R23{${\rm R}_{23}$}
\def\dS23{${\rm S}_{23}$}

\def\Zsun{${\rm Z}_{\odot}$}
\def\Msun{${\rm M}_{\odot}$}

\def\NIIOII{[\ion{N}{2}]/[\ion{O}{2}]}

\def\OH{$\log({\rm O/H})+12$}

\def\NIIOII{[\ion{N}{2}]/[\ion{O}{2}]}

\def\ratioS23{([\ion{S}{2}]~$\lambda \lambda$6717,31 +
[\ion{S}{3}]~$\lambda\lambda$9069,9532)/H$\beta$}
\def\NIIHa{[\ion{N}{2}]/H$\alpha$}

\def\SII{[{\ion{S}{2}}]}

\def\Hb{{H$\beta$}}

\def\O4363{[{\ion{O}{3}}]~$\lambda$4363}
\def\OIII{[{\ion{O}{3}}]}

\def\Ha{{H$\alpha$}}

\def\L60{L$_{60}$}

\def\D4{$D4000$}

\def\Te{T$_{\rm e}$}

\shorttitle{}
\shortauthors{}

\begin{document}

\title{SDSS~0809+1729:  Connections Between Extremely Metal Poor Galaxies and 
Gamma Ray Burst Hosts}


\author{Lisa J.\ Kewley\altaffilmark{1}}
\affil {University of Hawaii}
\affil{2680 Woodlawn Drive, Honolulu, HI 96822}
\email {kewley@ifa.hawaii.edu}
\altaffiltext{1}{Hubble Fellow}

\author{Warren R.\ Brown\altaffilmark{2},
	Margaret J.\ Geller,
	Scott J.\ Kenyon, \and
	Michael J.\ Kurtz}
\affil{Smithsonian Astrophysical Observatory}
\affil{60 Garden Street MS-20, Cambridge, MA 02138}
\altaffiltext{2}{Clay Fellow, Harvard-Smithsonian Center for Astrophysics}

\begin{abstract}
	We discuss the serendipitous discovery of an extremely metal poor
galaxy, SDSS~0809+1729, classified as a star in the Sloan Digital Sky Survey
(SDSS). The galaxy has a redshift $z = 0.0441$ and a $B$-band absolute
magnitude M$_B = -17.1$. With a metallicity of \OH $\sim 7.44$ or $\sim1/20$
solar, this galaxy is among the 10 most metal poor emission-line galaxies
known. SDSS~0809+1729 is a blue compact galaxy (BCG) with a stellar age of
$\sim 4.5$~Myr, a star formation rate of 0.18 \Msun yr$^{-1}$, and a large gas-phase
electron density ($\sim367$~cm$^{-3}$). Similar values of these parameters are
common among other extremely metal poor galaxies, including I~Zw~18.  
SDSS~0809+1729 is, however, unusual among BCGs because it lies in the same
region of the luminosity-metallicity diagram as the two lowest metallicity long
duration gamma ray burst (GRB) hosts.  For a given $B$-band luminosity, both
nearby GRB hosts and SDSS~0809+1729 have systematically lower metallicities than dwarf
irregulars, the majority of BCGs, and normal star-forming galaxies. Because the
star formation properties of SDSS~0809+1729 are similar to nearby long duration GRB
hosts, SDSS~0809+1729 may be a potential GRB host.  Identification of larger
samples of similar extremely metal poor objects may provide important insights
into the conditions required to produce long duration GRBs. \end{abstract}


\section{Introduction}
	Extremely metal poor local galaxies are key to understanding star
formation and enrichment in a nearly pristine interstellar medium (ISM).  
Metal poor galaxies provide important constraints on the pre-enrichment of the
ISM by previous episodes of star formation. Evolutionary scenarios proposed to
explain the low metallicities include metal loss from supernova-driven winds
\citep{Ferrara00,Recchi04}, dilution of the ISM metallicity by infall of
unenriched gas \citep{Koppen05}, or a star formation history dominated by short
bursts of star formation separated by long quiescent periods \citep{Searle72}.  
The latter scenario is currently favored \citep[see][for a review]{Kunth00},
but all of these mechanisms probably play some role.

	Intriguingly, metal poor galaxies have recently been linked to long
duration GRBs \citep{Stanek06,Wolf06,Fruchter06}. \citet{Sollerman05} compared
the host galaxy properties of three nearby GRBs with the properties of local BCGs. 
They found that the luminosity and star formation rates of the GRB hosts
are similar to those of local BCGs.  \citet{Fruchter06} demonstrate that long
duration GRBs are concentrated in the highest surface brightness regions of
their extremely blue hosts.

	\citet{Stanek06} show that five nearby GRB hosts have lower
metallicities than normal star-forming galaxies of similar luminosity.  
Because the GRB energy released decreases steeply with increasing host
metallicity, Stanek et al.\ propose an upper metallicity limit for cosmological
GRBs of $\sim 0.15$~\Zsun.  Strong Lyman-$\alpha$ detections in GRB hosts at
$z\gtrsim 2$ also support low host metallicities \citep{Fynbo03}.


	Extremely metal poor galaxies (XMPGs) are rare.  Fewer than 1\% of
dwarf galaxies have extremely low metallicities, defined as \OH$\leq 7.65$
\citep{Kunth00,Kniazev03}.  XMPGs, however, are much more common than known GRB
hosts.  Thus a larger sample of XMPGs is one route to understanding the
environment required for the formation of the long duration GRBs.  Most of the
known XMPGs are gas-rich BCGs with spectra dominated by emission lines
\citep[e.g.,][]{Kunth83,Thuan95}. Emission-lines provide an estimate of the
gas-phase metallicity from the \OIII~$\lambda 4363$ doublet.  This line is
particularly sensitive to the electron temperature of the gas and is strong in
metal poor galaxies where \HII~region cooling is minimal.

	Many investigators have searched for XMPGs with varying degrees of
success.  For more than three decades, the famous galaxy I Zw 18 was the most
metal poor galaxy known, with a metallicity of \OH$\sim7.17$ \citep{Searle72}.  
Recently, \citet{Izotov05} showed that the galaxy SBS 0335-052W has an even
lower oxygen abundance of \OH$\sim7.12$. Searches of SDSS spectra based on
emission-lines \citep{Izotov04,Kniazev03,Kniazev04,Papaderos06a,Izotov06b} have uncovered
few XMPGs, in agreement with previous searches of emission-line galaxies
\citep[e.g.,][]{Terlevich91,Masegosa94,VanZee00}.  Surveys that select on the
equivalent width of strong lines such as \OII\ appear to be more efficient at
finding XMPGs \citep{Ugryumov03}, but the number of known XMPGs remains small.

	Here we report the serendipitous discovery of a new, very blue XMPG,
SDSS~0809+1729.  This galaxy is at a redshift $z = 0.0441$ and has a
metallicity of \OH$\sim7.44$.  This metallicity places SDSS~0809+1729 among the
10 lowest metallicity emission-line galaxies known to date.

	We describe the observations and derived quantities in \S~\ref{obs}. We
compare the properties of our galaxy with those of other XMPGs in
\S~\ref{comp}.  We show that XMPGs and GRB hosts share similar spectral
properties in \S~\ref{GRB}.  We discuss our results and conclusions in
\S~\ref{Conclusions}.  Throughout this paper, we adopt the flat
$\Lambda$-dominated cosmology as measured by the WMAP experiment
\citep[$h=0.72$, $\Omega_{m}=0.29$;][]{Spergel03}).

\section{Observations and Derived Quantities}\label{obs}

	We discovered the low metallicity galaxy, SDSS~0809+1729, in the
hypervelocity star survey of \citet[][hereafter B06]{Brown06}.  B06 selected
candidate B-stars using SDSS $(u'-g')_0$, $(g'-r')_0$, and $(r'-i')_0$ colors.  
SDSS~0809+1729 is classified as a star in the SDSS, but spectroscopy shows that
it is a BCG with $(u'-g')_0=0.287$, $(g'-r')_0=-0.416$, and $(r'-i')_0=-0.058$
\citep[colors have been dereddened using][]{Schlegel98}. This dwarf galaxy is
at a redshift $ z = 0.0441$ and has an absolute magnitude M$_B = -17.1$.  
SDSS~0809+1729 has an observed $(B-R)_0 = -0.25$ and a rest frame $(B-R)_0 =
-0.27$, placing it in the blue tail of a large sample of nearby BCGs \citep[see
Figure 4,][]{Gil03}.

\begin{figure*}[!ht]
\plotone{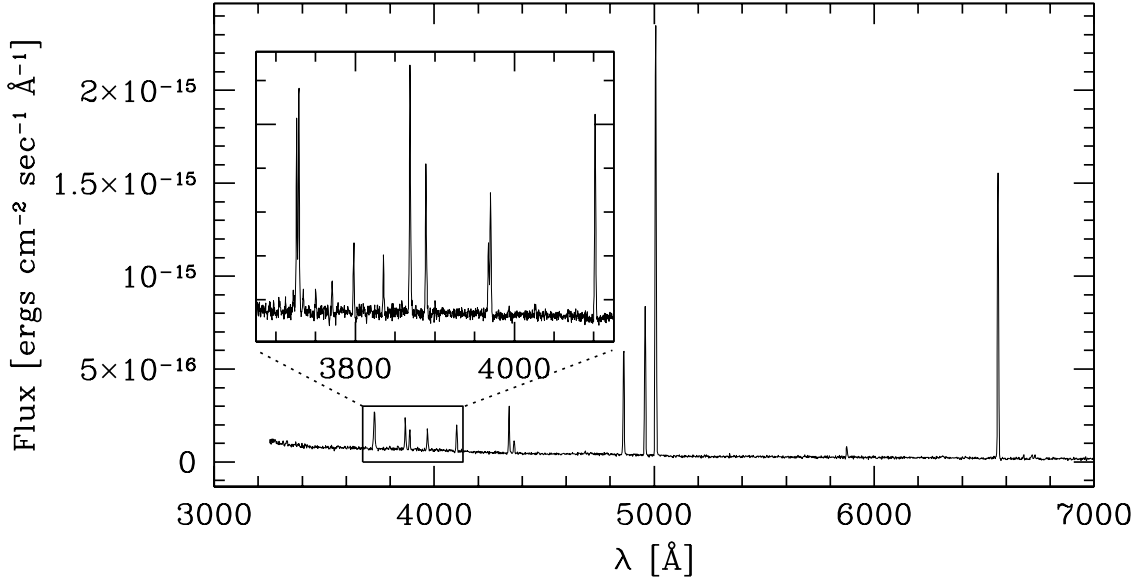}
\caption{\label{spec}
	MMT spectrum of SDSS~0809+1729 obtained with the 300 lines mm$^{-1}$
grating. SDSS~0809+1729 has a blue continuum and high \Ha\ and \OIII\
equivalent widths, indicating extremely low metallicity. The inset shows the
MMT spectrum obtained with the 832 line mm$^{-1}$ grating. [OII] is resolved
and the higher order Balmer lines are prominent in the inset spectrum.}
\end{figure*}

	We observed SDSS~0809+1729 on the Blue Channel spectrograph at the 6.5m
MMT telescope. In photometric conditions, we obtained a high resolution
spectrum with the 832 lines mm$^{-1}$ grating in second order, resulting in
wavelength coverage of 3650\AA\ to 4500\AA\ at a spectral resolution of 1.2\AA.
In non-photometric conditions, we obtained a lower resolution spectrum with
the 300 lines mm$^{-1}$ grating, resulting in broader wavelength coverage
(3400\AA\ to 8600\AA) at a resolution of 6.2 \AA.  
Figure~\ref{spec} shows both the high and low resolution spectra.

	We reduced the spectra using standard IRAF\footnote{
        IRAF is distributed by the National Optical Astronomy Observatories,
which are operated by the Association of Universities for Research in
Astronomy, Inc., under cooperative agreement with the National Science
Foundation.}
	spectral reduction tasks and flux calibrated the spectra using spectrophotometric
standards.  Flux calibration is accurate to $\sim$10\%.  We obtain relative
flux calibration for the non-photometric low resolution spectrum.  We fit the
optical emission-lines with gaussians using the {\it splot} task in IRAF.  The observed line fluxes and 
statistical errors are given in Table~\ref{flux_table}.  For the following analysis, we corrected the emission line fluxes for reddening using the Balmer decrement and the \citet{Cardelli89} 
(CCM) reddening curve.  We assumed an ${\rm R_{V}=A_{V}/{\rm E}(B\!-\!V)} = 3.1$ and an 
intrinsic H$\alpha$/H$\beta$ ratio of 2.85 (the Balmer decrement for case B 
recombination at T$=10^4$K and $n_{e} \sim 10^2 - 10^4 {\rm cm}^{-3}$;
Osterbrock~1989).   

\epsscale{1.0}

	The Balmer decrement of SDSS~0809+1729 is 2.76, consistent with zero
extinction, assuming an intrinsic \Ha/\Hb\ ratio of 2.86 \citep[the Balmer
decrement for case B recombination at T$=10^4$K and $n_{e} \sim 10^2 - 10^4
~{\rm cm}^{-3}$;][]{Osterbrock89}.

	We used the \SII~$\lambda 6717$/ \SII~$\lambda 6731$ line ratio in
conjunction with a 5 level model atom using the Mappings photoionization code
\citep{Sutherland93} to calculate an electron density of ${\rm n_e }
=367$~cm$^{-3}$ in the S$^{+}$ zone.  The \OII~$\lambda 3726$/\OII~$\lambda
3729$ ratio gives ${\rm n_e } =195$~cm$^{-3}$ in the O$^{+}$ zone.

	We derive the gas-phase oxygen abundance  following the procedure
outlined in \citet{Izotov06a}.  This procedure utilizes the
electron-temperature (\Te) calibrations of \citet{Aller84} and the atomic data
compiled by \citet{Stasinska05}.   Abundances are determined within the framework of 
the classical two-zone HII-region model \citep{Stasinska80}.
The ratio of the auroral \OIII~$\lambda 4363$
and \OIII~$\lambda \lambda 4959,5007$ emission-lines give an electron
temperature in the O$^{++}$ zone of ${\rm T_{e}}=19\times 10^4$~K assuming an
electron density of 195~cm$^{-3}$.  These electron temperatures are insensitive
to small variations in electron density; we obtain the same ${\rm T_{e}}$ with an electron density of 367~cm$^{-3}$.  The electron temperature of the 
O$^{+}$ zone is calculated assuming ${\rm T_{e}}({\rm O}^{+})=0.7\, {\rm T_{e}}({\rm O}^{++} ) + 0.3 $
\citep{Stasinska80}.  We calculate the metallicity in
the O$^{+}$ and O$^{++}$ zones assuming
	\begin{equation}
{\rm O/H}={\rm O}^{+}/{\rm H}^{+} + {\rm O}^{++}/{\rm H}^{+}
\end{equation}

The resulting metallicity for SDSS~0809+1729 is \OH$\sim 7.44$.  The uncertainty in the 
absolute O/H metallicity determination 
is $\sim 0.1$~dex. This intrinsic uncertainty is the dominant error in our metallicity
determination, and includes errors in the use of simplified \HII\ region models and
possible problems with electron temperature fluctuations \citep{Pagel97}.  Fortunately, these errors 
affect all \Te-based methods in a similar way and the error in relative metallicities derived using the same method is likely to be $<< 0.1$~dex.  

We use the \Te-based metallicities to facilitate comparisons with metal poor galaxies in the literature.   We also provide metallicity measurements using the theoretical strong-line diagnostics 
from \citet[][; hereafter KD02]{Kewley02a}, updated in \citet{Kobulnicky04}.
The KD02 method is based on the \NIIOII\ ratio for high metallicities (\OH $\gtrsim8.4$), and the (\OII $+$\OIII)$/{\rm H \beta}$ (\R23) diagnostic for low metallicities (\OH $\lesssim8.4$).  The majority of galaxies analysed in this paper have low metallicities (as indicated by their \NIIOII\ or \NIIHa\ ratios) and we use the \citet{Kobulnicky04} re-calibration of the KD02 \R23\ diagnostic.  The absolute error in the strong-line metallicities is $\sim 0.15$~dex.  There is a well-known discrepancy of $\sim$0.4~dex between metallicities calculated using theoretical strong-line diagnostics and diagnostics based on electron-temperature measurements.  The cause of this offset is unknown \citep[see][for a discussion]{Garnett04,Garnett04b,Stasinska05}.  Until this discrepancy is resolved, 
absolute metallicities derived using any method should be treated with caution.  
Fortunately, the difference between the strong-line and \Te\ diagnostics is systematic.  Therefore
the error in relative metallicities derived using {\it the same} method, 
regardless of the method, is $<< 0.1$~dex \citep{Ellison05,Kewley06}.  
We therefore use metallicities based on the \Te-method, or converted into the \Te\ method
only for relative comparisons between galaxies.

	The ionization parameter of a gas $q$ is a measure of the level of
ionization that a radiation field can drive through a nebula. The ionization
parameter is defined as the number of hydrogen ionizing photons passing through
a unit area per second per unit hydrogen number density.  We calculate the ionization 
parameter using the
iterative prescription outlined in \citet{Kewley02a}.  The ionization parameter
for SDSS~0809+1729 is $q= 1.5\times10^{8}$~cm s$^{-1}$, a value that is
extremely high relative to normal star-forming galaxies
\citep[e.g.][]{Dopita06}.  Such a high ionization parameter indicates that the
stellar EUV radiation field and the electron temperature in SDSS~0809+1729 are
stronger than in most star-forming galaxies.  In Section~\ref{comp} we
investigate the star formation properties of SDSS~0809+1729 in more detail.

\section{Comparison with other Extremely Low Metallicity Galaxies}\label{comp}

	We compare the spectral properties of SDSS~0809+1729 with those of
other XMPGs for which both (1) ${\rm T_{e}}$-based metallicities are available,
and (2) \Ha\ photometry or global spectra are available. 

	Because SDSS~0809+1729 is stellar in appearance, our spectra provide
global estimates of galaxy properties.  We note that the metallicities of the
more extended low metallicity galaxies in the literature often lack global
estimates of metallicity or other parameters.  For example, the metallicity of
7.17 cited for J0113+0052 \citep{Izotov06b} was obtained from spectra of an
\HII\ region on the outskirts of the low surface brightness galaxy UGC~772. A
global metallicity estimate of UGC~772 may differ from the \HII\ region value.
Where possible, we compare the properties of SDSS~0809+1729 only with other low
metallicity galaxies that have global estimates.

\subsection{Stellar Population Age}

	The detailed star formation histories of XMPGs are difficult to
measure; most XMPGs, like SDSS~0809+1729, are too far away or too compact to
resolve into individual stars, even with HST.  XMPGs remain a puzzle because
they may be pristine galaxies undergoing their first burst of star formation or
they may contain an older stellar population from previous episodes of star
formation.  \citet{Izotov99} calculate C/O and N/O abundances for a sample of
54 BCGs.  They find very little scatter in the C/O and N/O ratios in BCGs with
\OH$\leq 7.6$.  Izotov \& Thuan suggest that such low scatter rules out
time-delayed production of C and primary N and that extremely low metallicity
BCGs are therefore undergoing their first burst of star formation.  The
fraction of neutral gas in XMPGs ($\sim 99$\% of all baryonic mass) gives
further evidence against an old stellar population
\citep{VanZee98,Pustilnik01}.  On the other hand, metal-poor galaxies may have
extremely faint old stellar populations if their previous episodes of star
formation occurred at very low star formation rates \citep{Legrand00}.  HST
images of the nearest XMPGs provide some evidence for this picture.
Color-magnitude diagrams derived from these images provide, for example, ages
for I Zw 18 which range from $\sim 500$~Myr \citep{Izotov04b} to $\sim 1$~Gyr
\citep{Aloisi99}; the stellar age from similar data for the metal poor BCG SBS
1415+437 is $\sim 1.3$~Gyr \citep{Aloisi05}.

	Although we cannot estimate the age of an old stellar population (if
any) in SDSS~0809+1729, stellar population synthesis models provide an estimate
of the age of the young stellar population given a metallicity, initial mass
function (IMF), and star formation history prescription.  The Balmer line
equivalent widths (EWs) are strongly correlated with the age of the stellar
population \citep{Schaerer98,Gonzalez99}.  Assuming a Salpeter IMF and an
instantaneous burst of star formation, the \Hb\ equivalent width of 88.25\AA\
corresponds to a stellar age of $\sim 4.5$~Myr using the $Z=0.001$
(\OH$\sim7.6$)  models of \citet{Schaerer98}.
  
\begin{figure*}[!t]
\plotone{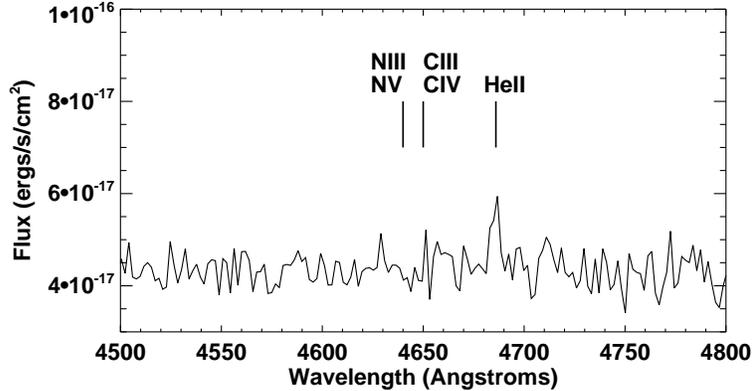}
\caption{\label{WR_spec}
	Spectrum of SDSS~0809+1729 showing the positions of expected Wolf-Rayet
features.  The N and C Wolf-Rayet features are undetected.}
\end{figure*}

	An age of 4.5~Myr corresponds to the end of the Wolf-Rayet phase for a
metallicity of $Z=0.001$.  Figure~\ref{WR_spec} shows the spectrum of
SDSS~0809+1729 between 4500\AA-4800\AA\ with the positions of expected
Wolf-Rayet features marked.  The HeII~$\lambda 4686$ line is clearly detected
in the spectrum, but the carbon and nitrogen Wolf-Rayet features are absent.  
These carbon and nitrogen lines typically blend to produce the "blue bump"
frequently seen in galaxies during the Wolf-Rayet phase.

	The lack of prominent Wolf-Rayet features supports a stellar age of
$\sim4.5$~Myr. Narrow HeII emission is associated with nebular emission rather
than with the broader stellar emission from Wolf-Rayet stars.  It is
interesting to note that our log(HeII/\Hb) ratio (-1.62)  exceeds the value for
pure nebular emission according to the instantaneous burst models of
\citet[][]{Schaerer98}.  We also note that the lack of Wolf-Rayet features does
not necessarily indicate that the Wolf-Rayet phase is over. \citet{Crowther06}
showed that theoretical models of Wolf-Rayet stars in extremely low metallicity
environments produce weaker Wolf-Rayet features by a factor of $3-6$ than in
galaxies at metallicities similar to the LMC.

\epsscale{0.7}

	Other XMPGs have young stellar ages similar to SDSS~0809+1729
(Table~\ref{LowZ_table}).  All of the XMPGs in our comparison sample have
stellar ages within the narrow range of 3-4.5~Myr, according to the $z=0.001$
Schaerer \& Vacca models.  This age range corresponds to the middle to end of
the Wolf-Rayet phase.

\subsection{Star Formation Rates}

	The derivation of star formation rates (SFRs) for XMPGs is non-trivial.  
The commonly used \Ha-based SFR calibrations such as \citet{Hunter86} and
\citet{Kennicutt98} were developed for ``normal'' dwarf irregular and spiral
galaxies, respectively.  These traditional calibrations were derived using
stellar population synthesis models at solar metallicities, and are not
applicable to metal-poor galaxies where the SFR can vary on short timescales.
\citet{Weilbacher01} showed that (1) very low metallicity, (2) a time delay between the
onset of star formation and maximum H$_\alpha$-luminosity, and (3) a varying
stellar IMF yields SFRs differing by factors of 3-100 from results derived with
traditional calibrations.

	\citet[][hereafter B05]{Bicker05} use new stellar population synthesis
models to calculate \Ha\ SFR calibrations as a function of metallicity.  They
show that for galaxies at low metallicities \OH$\leq 8.2$ (Z=0.004 in B05) the
constants that convert L(\Ha) into SFRs differ substantially from the solar
value.  B05 provide calibrations for 5 values of metallicity because they
restrict the population synthesis models to the specific metallicities for
which stellar evolution tracks are available.  The two lowest metallicities of
$Z=0.0004$ and 0.004 correspond to \OH$=7.23$ and 8.23 respectively.  Because
SDSS~0809+1729 and most of our comparison galaxies have metallicities between
these values, we fit a polynomial to the inverse of the constants in B06 to
obtain a calibration for arbitrary metallicity values.  The \Ha\ calibration is
then:
	\begin{equation}
{\rm SFR(H\alpha)} ({\rm M}_{\odot}/{\rm yr}) =
1\times10^{-41}\,(-0.0954+0.0644*Z_m)\,   {\rm L(H\alpha) (ergs/s)}
\label{SFR}
\end{equation}
\noindent where $Z_m$ is the metallicity in units of \OH.


	Application of this calibration to the galaxies in
Table~\ref{LowZ_table} assumes that XMPGs have
similar star formation histories.  We provide two SFR estimates for each galaxy, based on either
 \Te-based or KD02 strong-line metallicities.  The difference between the two metallicity-corrected 
SFR estimates is minimal for the low SFRs in XMPGs.  For SDSS~0809+1729, we use H$\delta$ from
our high resolution spectrum to calculate a rest-frame \Ha\ luminosity of
$4.07\times10^{40}$~ergs s$^{-1}$.  If we use \Ha\ from our lower resolution spectrum
obtained under non-photometric conditions, we obtain a similar rest-frame \Ha\ luminosity
of $4.39\times10^{40}$~ergs s$^{-1}$.  Applying equation~\ref{SFR} gives an
SFR(\Ha) of 0.18~\Msun yr$^{-1}$.  This SFR is large compared to most of the
other XMPGs; SDSS~0809+1729 has a SFR that is 10 times larger than the SFR in I
Zw 18.

\section{Comparison with GRB Hosts}\label{GRB}

	XMPGs can be hosts of long-duration GRBs \citep{Stanek06,Fruchter06}.  
\citet{Fruchter06} outline some of the physics underlying the connection
between long duration GRBs and their low metallicity hosts. They argue that the
progenitors of the GRBs are stars massive enough to form a black hole
surrounded by an accretion disk when the core collapses. Massive metal-rich
stars lose mass in stellar winds and thus may not retain enough mass to collapse
to black holes. Massive metal-poor stars lack the opacity to support a stellar
wind. They lose little mass and, unlike their metal-rich counterparts, are not
surrounded by extended HII regions. The supernova ejecta for metal-poor massive
stars thus do not sweep up H and He in the surrounding wind and can produce the
anomalous Type 1C supernovae associated with some nearby long duration GRBs.  
This plausibility argument motivates further identification and study of
larger samples of XMPGs as potential long duration GRB hosts.

	In this section we demonstrate that SDSS~0809+1729 has properties
remarkably similar to the hosts of two of the long duration GRB hosts, GRB
030329 and GRB060218.  We include a broader comparison of SDSS~0809+1729 with
other XMPGs and with known long duration GRB hosts. We examine extinction,
SFRs, and position with respect to luminosity-metallicity relations.

\subsection{Extinction}

	In Table~\ref{GRB_table} we list the \Ha/\Hb\ ratios from the
literature for the five GRB hosts in \citet{Stanek06}. The spectra of GRB hosts
generally suffer very little extinction, similar to the spectra of XMPGs.  We
note however that a small fraction of GRB hosts are detected in the infrared
and sub-mm, indicating that some GRB hosts contain dust
\citep{Frail02,Tanvir04,LeFloch06}.

\subsection{Star Formation Rates and Star Formation History}

	The SFRs in a few of the nearest GRB hosts have been constrained from
X-ray data \citep{Watson04},from radio and sub-millimeter measurements
\citep{Berger03}, and from optical \citep{Sollerman05,Gorosabel05}, and UV
observations \citep{Fynbo03,Christensen04,Jakobsson05}.  These studies usually
yield low SFRs that span about two decades from 0.1-10~\Msun yr$^{-1}$.

	We calculate the GRB host SFRs using the \Ha\ luminosity to facilitate
comparisons with the SFR(\Ha) of the extremely low metallicity galaxies.  The
GRB hosts have low metallicities (\OH$\leq 8.0$) and therefore they are subject
to the metallicity problems outlined by \citet{Weilbacher01} and B05.  We therefore apply
the B05 SFR calibration parameterized in equation~\ref{SFR}, taking into account the 
low GRB host metallicities.  The resulting
SFRs are given in Table~\ref{GRB_table}.

	The SFR(\Ha) for GRB hosts ranges between 0.03-9 \Msun yr$^{-1}$.  The
majority (4/5) of the GRB hosts have SFRs less than 0.2~\Msun yr$^{-1}$.  
These SFRs are remarkably similar to the SFRs in the extremely low metallicity
galaxies.  In particular, SDSS~0809+1729 has an SFR that is very close to three
of the GRB hosts (0.18~\Msun yr$^{-1}$ c.f. 0.18,0.16, and 0.12~\Msun
yr$^{-1}$).

	\citet{Sollerman05} found that the luminosity and SFRs of the GRB
hosts are similar to those of local BCGs.  Our GRB and low metallicity
galaxy SFRs extend the Sollerman et al. result to lower metallicity
galaxies.

	For the two GRB hosts where \Hb\ EWs are published, the age of the
young stellar population is 4.5~Myr, and 9~Myr, indicating populations at the
end of the Wolf-Rayet phase.  Wolf-Rayet features have been detected in some
extremely low metallicity galaxies \citep{Izotov01,Izotov97,Legrand97}, and
have recently been detected in GRB hosts \citep{Hammer06}. Although we do not
see Wolf-Rayet features in SDSS~0809+1729, we derive a population age of 4.5~Myr
and conclude that the populations is also at or near the end of the Wolf-Rayet
phase.

\subsection{Luminosity-Metallicity Relation}\label{SDSS}

	The luminosity-metallicity relation is an important tool for
investigating the chemical enrichment and mass-loss in galaxies.
\citet{Rubin84} provided the first evidence that metallicity is correlated with
luminosity in disk galaxies.  Larger samples of nearby disk galaxies solidified
the luminosity-metallicity relation \citep{Bothun84,Wyse85, Skillman89,Vila92,
Zaritsky94,Garnett02,Baldry02,Lamareille04,Tremonti04}.  A strong
luminosity-metallicity relation also exists for dwarf irregular (dIrr) galaxies
\citep{Lequeux79,Skillman89,Richer95,Pilyugin01,Garnett02,Lee04,Henry06} and
BCGs \citep{Campos93,Shi05}.  The luminosity-metallicity relation is generally
attributed to the more fundamental mass-metallicity relation that holds for
star-forming disk galaxies \citep{Tremonti04}.  Low mass galaxies have larger
neutral gas fractions than more massive galaxies
\citep{McGaugh97,Bell00,Boselli01}.  Selective loss of heavy elements from low
mass galaxies in supernova-driven outflows also contribute to the relation
\citep{Garnett02,Tremonti04}.

\begin{figure*}[!ht]
\plotone{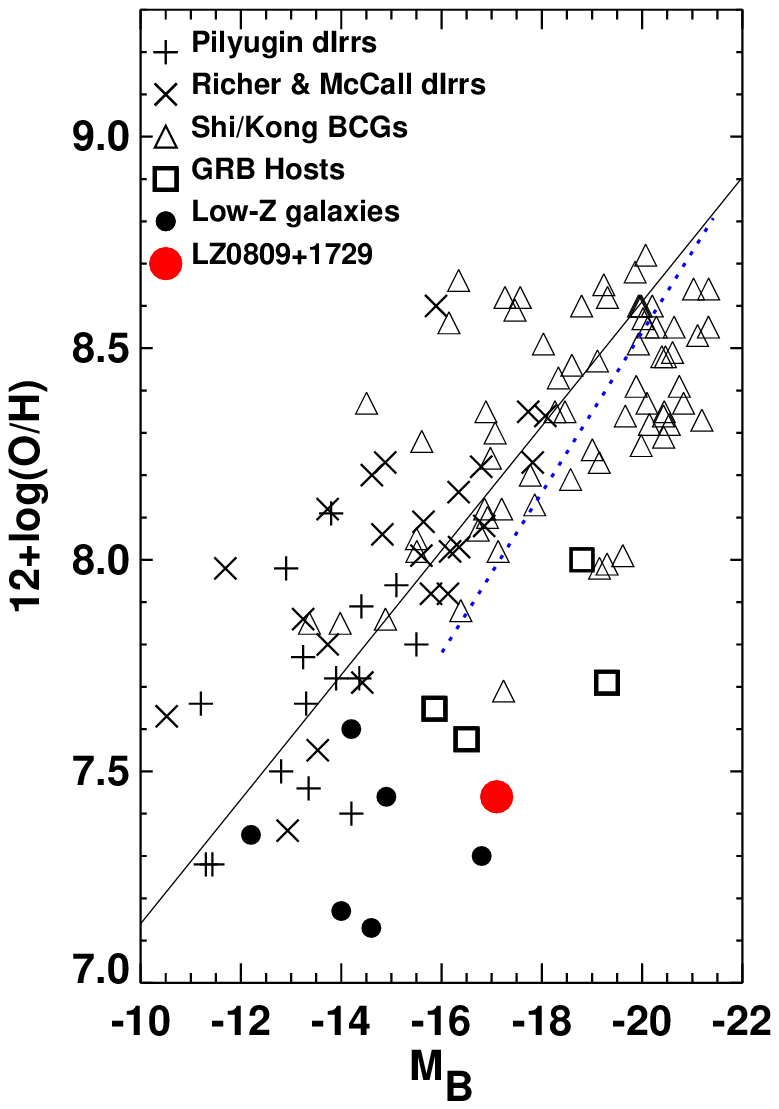}
\caption{\label{LZplot}
	$B$-band absolute magnitude versus gas-phase oxygen abundance for dIrrs
\citep[crosses,][]{Richer95} \citep[plusses,]{Pilyugin01}, BCGs \citep[open
triangles,][]{Kong04,Shi05}, GRB hosts \citep[open squares,][]{Stanek06}, and
extremely metal poor BCGs with ${\rm T_{e}}$-based metallicities and \Ha\
photometry or global spectra available (solid circles).
	The solid line gives the mean LZ relation for dIrrs from
\citet{Richer95} while the dotted line gives the SDSS relation for star-forming
galaxies in the SDSS DR4 from Kewley \& Ellison (2006, in prep).
	Our newly discovered XMPG, SDSS~0809+1729, is shown as a large solid
red circle. SDSS~0809+1729 lies in the same region of the diagram as the two 
lowest metallicity GRB hosts.  Note that the metallicities used in this figure, including the mean LZ relations are either (1) calculated using 
the \Te-method, or (2) converted into \Te\ metallicities using Kewley \& Ellison (2006).  The \Te-based metallicities are lower than strong-line metallicity estimates such as \citet{Kewley02a} by $\sim0.4$~dex.}
\end{figure*}

	In Figure~\ref{LZplot}, we show where SDSS~0809+1729 (large solid
circle) lies relative to the luminosity-metallicity relations of normal dIrrs
\citep{Richer95,Pilyugin01} (crosses, plusses), BCGs \citep{Kong04,Shi05}, and
star-forming galaxies from SDSS \citep{Tremonti04} (dotted line).  We show the
positions of the GRB hosts from \citet{Stanek06} (squares) and the lowest
metallicity galaxies from Table~\ref{LowZ_table} (small filled circles).  The
properties of the GRB hosts are listed in Table~\ref{GRB_table}.  Where the
\OIII~$\lambda 4363$ line is available, we have recalculated the metallicities
for the GRB hosts using the \Te-method.  For cases where only a limited range
of strong emission-lines is available, we apply the metallicity conversions of
\citet{Kewley06} to place the strong-line (KD02) metallicities on the same 
\Te-based scale.   The KD02-\Te\ conversion is valid for KD02 metallicities $12+\log({\rm O/H)_{KD02}}<8.2$ and is 
given by:

\begin{equation}
12+\log({\rm O/H})_{\rm T_{e}} = \frac{(12+\log({\rm O/H})_{\rm KD02}) - 0.613}{0.975} \label{KD02-Te}
\end{equation}

\epsscale{0.7}

	SDSS~0809+1729 lies remarkably close to the two lowest metallicity GRB
hosts, GRB~060218 and GRB~030329.  The XMPGs (including SDSS~0809+1729) and the
GRB hosts lie offset from the dIrr and SDSS luminosity-metallicity relations;
at a fixed luminosity, the XMPGs and GRB hosts are more metal poor than dIrr or
SDSS galaxies.  It is not clear whether this offset is a metallicity or a
luminosity effect (or both).  Either (1)  GRB hosts and XMPGs are less enriched
(by 0.2-0.4~dex) than dIrrs and SDSS galaxies for their luminosity, or (2) GRB
hosts and XMPGs are over-luminous in the $B$-band (by 2-4 magnitudes) for their
metallicities relative to dIrrs and SDSS galaxies.  The $B$-band luminosities
of nearby XMPGs include a significant contribution from emission-lines.  
However, the contribution of emission-lines to the total blue luminosity is
33\% for SDSS~0809+1729, sufficient to raise the luminosity by only $\sim$0.3
magnitudes. Stellar mass estimates derived from deep near infrared photometry
may help resolve this metallicity versus luminosity issue.     Nevertheless, Figure~\ref{LZplot} indicates that studies of GRB hosts should not assume that GRB hosts lie along the normal luminosity-metallicity relation. 
  
	GRB hosts at higher redshifts also show an offset from the normal
luminosity-metallicity relation; 5/7 GRB hosts at $0.4<z<1$ lie below the mean
luminosity-metallicity relation for normal star forming galaxies at $0.4<z<1$
\citep[figure 3 of][]{Savaglio06}.  The offset at $0.4<z<1$ between GRB hosts
and normal star-forming galaxies is smaller (0.5~mag in luminosity or 0.2~dex
in metallicity) than the offset apparent in Figure~\ref{LZplot}. However,
this difference may result from the large errors in the metallicity estimates
at $0.4<z<1$ \citep{Kobulnicky04}.

\section{Conclusions}\label{Conclusions}

	We report the serendipitous discovery of an extremely metal poor
galaxy, SDSS~0809+1729, classified as a stellar point source in the SDSS
catalog.  This galaxy has a metallicity of \OH$\sim7.44$, or $\sim 1/20$~solar,
placing it in the group of ten lowest metallicity emission-line galaxies.  We
compare the spectral properties of SDSS~0809+1729 with the spectral properties
of seven known XMPGs, including I Zw 18.  We apply stellar population synthesis
models to derive stellar ages and we calculate SFRs, taking into account the
effect of metallicity.  The young stellar population in SDSS~0809+1729 has an
age of $\sim 4.5$~Myr, corresponding to the end of the Wolf-Rayet phase.  The
star formation rate in SDSS~0809+1729 ($\sim0.18$~\Msun yr$^{-1}$) is ten times
larger than the star formation rate in I Zw18.

	We compare the properties of SDSS~0809+1729 and the other XMPGs
with the properties of five nearby GRB hosts for which spectral data is available.  We 
find that:

\begin{itemize}
	\item XMPGs and GRB hosts lie in a similar region of the
luminosity-metallicity diagram: both types of objects are systematically offset
from the luminosity-metallicity relation defined by dIrrs, normal BCGs, and
normal star-forming galaxies.  For a given luminosity, XMPGs and GRB hosts are 
less enriched than dIrrs, normal BCGs, and normal star-forming galaxies.  

	\item XMPGs and GRB hosts share similar SFRs and extinction levels,
strengthening the link between GRB hosts and XMPGs.

	\item Our new XMPG, SDSS~0809+1729, has nearly identical properties as
the lowest metallicity GRB host, GRB~030329.  The \Te\ metallicities of the two
galaxies agree to within $\sim 0.1$~dex (the estimated error in the metallicity
diagnostics is $\sim 0.1$~dex), and the SFRs agree to within 0.06~\Msun
yr$^{-1}$, or 50\%.  While electron density measurements are unavailable for
GRB hosts with luminosities similar to SDSS~0809+1729, it is interesting that the
luminous host galaxy of GRB~031203 has an electron density (and stellar age)
very similar to the one we measure for SDSS~0809+1729.

\end{itemize}

	The similarity between XMPGs and long duration GRB hosts is remarkable.  
Our results suggest that SDSS~0809+1729 and other XMPGs like it are potential 
hosts for long duration GRBs.

	Identification and analysis of larger samples of XMPGs may provide
important insight into the environment where long duration GRBs occur.  The
serendipitous discovery of an XMPG using SDSS colors suggests that a similar
color-selection method may be useful for revealing new XMPGs. Some of these
objects have been missed because, like SDSS~0809+1729, they are
indistinguishable from very blue stars.  We will report soon on a more
extensive survey to identify these objects (Brown et al.\ 2006, in prep.).

\acknowledgments
We thank the anonymous referee for helpful comments, and we thank M. Modjaz, K. Stanek, and D. Sanders for useful discussions.  L. J. Kewley is supported by a Hubble Fellowship. W. R. Brown is supported by a Clay Fellowship. This research has made use of NASA's
Astrophysics Data System Bibliographic Services. We thank the Smithsonian
Institution for partial support of this research.

\clearpage
\bibliography{library}

\newpage

\begin{deluxetable}{ll}
\tabletypesize{\scriptsize}
\tablecaption{Line intensities for SDSS 0809+1729
\label{flux_table}}
\tablehead{$\lambda_{0}$ Ion (\AA) 
& F($\lambda$)/F(H$\beta$)\tablenotemark{a} }
\startdata
3726~[OII] & 0.20$\pm$ 0.02\\
3729~[OII] & 0.25$\pm$ 0.02\\
3750~H12 & 0.022$\pm$ 0.006\\ 
3771~H11 & 0.033 $\pm$ 0.005\\ 
3798~H10 & 0.066 $\pm$ 0.009 \\ 
3835~H9 & 0.055 $\pm$ 0.008\\ 
3869~[NeIII] & 0.29$\pm$ 0.02\\
3889~H8$+$HeI & 0.17 $\pm$ 0.01\\ 
3967~[NeIII] & 0.079$\pm$ 0.007\\
3970~H7 & 0.138$\pm$ 0.009\\
4026~HeI & 0.013 $\pm$ 0.003\\ 
4101~H$\delta$ & 0.24$\pm$ 0.01\\
4340~H$\gamma$ & 0.415  $\pm$ 0.009\\ 
4363~[OIII]  & 0.129 $\pm$ 0.006\\
4686~HeII & 0.024$\pm$ 0.006\\
4681~H$\beta$ & 1.00 $\pm$ 0.01\\
4959~[OIII] & 1.37$\pm$ 0.02\\
5007~[OIII] & 4.03$\pm$ 0.04\\
5876~HeI & 0.105$\pm$ 0.007\\
6563~H$\alpha$ & 2.76$\pm$ 0.03\\
6678~HeI & 0.034$\pm$ 0.006\\
6717~[SII] & 0.042$\pm$ 0.007\\
6731~[SII] & 0.038$\pm$ 0.007\\
7065~HeI & 0.035$\pm$ 0.006\\
7136~[ArIII] & 0.025$\pm$ 0.07\\  \\
F(H$\beta$)\tablenotemark{b} & 33.0 $\pm$ 3.3 \\
EW(H$\beta$) & 88.3 $\pm$ 0.8\\ 
\enddata
\tablenotetext{a}{Errors in fluxes relative to H$\beta$ include poisson statistical errors for both F($\lambda$) and F(H$\beta$)}.
\tablenotetext{b}{F(H$\beta$) in units of $1\times10^{-16}$~ergs/s/cm$^{2}$.  Error includes both statistical errors and 10\% error in the absolute flux calibration.}
\end{deluxetable}


\clearpage

\begin{landscape}
\begin{deluxetable}{lllllllllllll}
\tabletypesize{\scriptsize}
\tablewidth{20cm}
\tablecaption{Comparison between spectral properties of extremely low
metallicity galaxies
\label{LowZ_table}}
\tablehead{ID
& Hubble
& $v$
 & \multicolumn{2}{c}{log(O/H)$+12$}
 & M$_{\rm B}$
& EW(\Hb)
& Stellar
& L(\Ha)
& \multicolumn{2}{c}{SFR (\Msun/yr)}
& $N_e$(SII)
& \Ha/\Hb \\
& Type
& (km s$^{-1}$)
& ${\rm T_{e}}$
& KD02
&
&
& Age (Myr)
& (ergs s$^{-1}$)
& ${\rm T_{e}}$
& KD02
& (cm$^{-2}$)
&}
\startdata
 SBS0335-052W\tablenotemark{a}        & dIrr,BCG  & 4017    & 7.13 & 7.5  & 
 -14.6  & 109        & 4            &  $5.92\times 10^{39}$\tablenotemark{g}  & 0.021 & 0.022 &
 330 & 2.74 \\

 I Zw 18\tablenotemark{b}                           & BCG        &
 751    & 7.17 & 7.6 & -14.0  & 104.5    & 4           &  $4.92\times
 10^{39}$\tablenotemark{g} & 0.018 & 0.019 & 104.7 & 2.89 \\


 SBS0335-052E\tablenotemark{c}           & BCG        & 4057  & 7.30 & 7.5 & 
 -16.8    & 89-382  & 3-4.5      &  $1.35\times10^{41}$\tablenotemark{g}  & 0.51 & 0.53  &           & 2.74 \\

HS0822+3542\tablenotemark{e}              & BCG   & 732      & 7.35 &  7.6 & 
-12.2 &  292        & 3  & $9.82\times10^{38}$\tablenotemark{g}  & 0.004 & 0.004 & \nodata & 2.74 \\

HS2134+0400\tablenotemark{d}              & BCG  & 5070  & 7.44 & 7.8  & 
-14.9   &  214        & 3            & $8.29\times 10^{39}$\tablenotemark{g}  & 0.031 & 0.033 &
236    & 3.00 \\

SDSS~0809+1729                                              & BCG   &
13232 & 7.44 & 7.8 &  -17.1  & 88.3     & 4.5         & $4.64\times10^{40}$
& 0.18 & 0.19 & 367            & 2.76 \\

SBS 1415+437\tablenotemark{f}             & BCG   & 609      & 7.60  & 7.9  &
-14.2 & 134,166  & 4          &   $4.59\times 10^{39}$\tablenotemark{g}  & 0.018 & 0.019 & 60,90     & 2.71,2.81\\
\enddata
\tablenotetext{a}{References for
data:\citet{Pustilnik04,Izotov05,Papaderos06}}
\tablenotetext{b}{Reference for data:\citet{Kong02,Gil03}}
\tablenotetext{c}{References for data:\citet{Melnick92,Hunt01,Izotov01,Pustilnik04,Papaderos06}}
\tablenotetext{d}{References for data:\citet{Pustilnik06,Gil03}; H$\alpha$ is not a global measurement}
\tablenotetext{e}{References for data:\citet{Kniazev00,Gil03,Hunter04}}
\tablenotetext{f}{References for data:\citet{Gil03,Guseva03b}}
\tablenotetext{g}{We calculated L(H$\alpha$) using the observed H$\alpha$ fluxes from references (a-f); L(H$\alpha$) has been converted to rest-frame and corrected L(H$\alpha$) for extinction using the Balmer Decrement and the \citet{Cardelli89} reddening curve.}
\end{deluxetable}
\clearpage
\end{landscape}

\clearpage

\begin{deluxetable}{llllllllllll}
\tabletypesize{\scriptsize}
\tablecaption{Comparison between spectral properties of GRB hosts and
SDSS~0809+1729
\label{GRB_table}}
\tablehead{ID
& $v$
 & \multicolumn{2}{c}{log(O/H)$+12$}
 & M$_{\rm B}$
& EW(\Hb)
& Stellar
& L(\Ha)
& \multicolumn{2}{c}{SFR  (\Msun)}
& $N_e$(SII)
& \Ha/\Hb \\
& (km s$^{-1}$)
& T$_{\rm e}$
& KD02
&
&
& Age (Myr)
& (ergs s$^{-1}$)
& T$_{\rm e}$
& KD02
& (cm$^{-2}$)
& }
\startdata
GRB~980425\tablenotemark{a}   & 2549   & \nodata & 8.5 & -17.65  & \nodata  &
\nodata & $4.14 \times 10^{40}$\tablenotemark{h} & 0.18 & 0.19   & \nodata & 2.45 \\

GRB~031203\tablenotemark{b}   & 31629 & 7.8\tablenotemark{g} & 8.1  & -19.3   &  90.72   &
4.5       & $2.17\times10^{42}$\tablenotemark{h} & 8.81 & 9.23  & 300 & 8.94 \\

GRB~020903\tablenotemark{c}    &  74950 & 8.0\tablenotemark{f}  & 8.3  & -18.8 & 36.08 &
9            & $3.76\times10^{40}$\tablenotemark{i} & 0.16  & 0.16   & \nodata &  3.26     \\

GRB~060218\tablenotemark{d}    & 10046  & 7.6\tablenotemark{g} & 8.1  & -15.86 &  \nodata &
\nodata   & $8.17\times10^{39}$\tablenotemark{h}     & 0.03  & 0.03 & \nodata & 3.0    \\

GRB~030329\tablenotemark{e}    & 50516 & 7.5\tablenotemark{g} & 8.0 & -16.5    & \nodata  &
\nodata  & $3.00\times10^{40}$\tablenotemark{h}     &  0.12 & 0.13   & \nodata  &  2.77 \\

SDSS~0809+1729                                   & 13232 & 7.44  & 7.8 &
-17.1    & 88.3     & 4.5         & $4.64\times10^{40}$ & 0.18 & 0.19  &  367 &
2.76 \\
\enddata
\tablenotetext{a}{References for data:\citet{Sollerman05,Stanek06}}
\tablenotetext{b}{Reference for data:\citet{Prochaska04,Sollerman05,Stanek06}}
\tablenotetext{c}{References for data:\citet{Stanek06,Bersier06,Hammer06}}
\tablenotetext{d}{References for data:\citet{Stanek06,Modjaz06}}
\tablenotetext{e}{References for data:\citet{Gorosabel05,Sollerman05,Stanek06}}
\tablenotetext{f}{Metallicity is calculated using the ${\rm T_{e}}$ method and the [OIII]~4363 line in \citet{Hammer06}}
\tablenotetext{g}{${\rm T_{e}}$ metallicity is calculated using the KD02-${\rm T_{e}}$ conversion from \citet{Kewley06} (equation~\ref{KD02-Te})}
\tablenotetext{h}{We calculated L(H$\alpha$) using the observed H$\alpha$ fluxes from references (a-f); L(H$\alpha$) has been converted to rest-frame and corrected L(H$\alpha$) for extinction using the Balmer Decrement and the \citet{Cardelli89} reddening curve.}
\end{deluxetable}

\end{document}